\documentclass[letterpaper,11pt]{article}

\usepackage{amsmath,amssymb}
\usepackage{eepic,epic}
\usepackage{epsfig}
\usepackage{graphicx}
\usepackage{color}

\usepackage{comment}
\usepackage{tabularx}
\usepackage{booktabs} 
\usepackage{pifont}
\newcommand{\score}{\mbox{\it score}}
\newcommand{\p}{{\rm P}}
\newcommand{\np}{{\rm NP}}
\newcommand{\electionsystem}{\ensuremath{\cal E}}\newcommand{\cale}{\ensuremath{\cal E}}
\newtheorem{theorem}{Theorem}[section]

\newcommand\qedblob{\ding{113}}
\def\literalqed{{\ \nolinebreak\hfill\mbox{\qedblob\quad}}}

\newenvironment{proofs}{\noindent{\bf Proof.}\hspace*{1em}}{\literalqed\bigskip}

\newcommand{\btwo}[2]{\scriptsize\begin{bmatrix}\textrm{\rm{}#1}\\\textrm{\rm{}#2}\end{bmatrix}}
\newcommand{\bthree}[3]{\scriptsize\begin{bmatrix}\textrm{\rm{}#1}\\\textrm{\rm{}#2}\\\textrm{\rm{}#3}\end{bmatrix}}

\newcommand{\EP}[3]{
\begin{center}
{\small 
\begin{tabularx}{0.98\columnwidth}{ll}
\toprule
\multicolumn{2}{c}{\textsc{#1}} \\
\midrule
{\bf Given:}\hspace*{-1em}   & \parbox[t]{0.8\columnwidth}{#2\vspace*{1mm}}  \\
{\bf Quest.:}\hspace*{-1em}& \parbox[t]{0.8\columnwidth}{#3\vspace*{.5mm}} \\ 
\bottomrule
\end{tabularx}
}
\end{center}
}

\let\shortcite\cite

\begin{document}
%
\title{More Natural Models of\\Electoral Control 
by Partition\thanks{Supported in part by grants       
DFG-ER-738/\{1-1,2-1\} and                                                   
NSF-CCF-\{0915792,1101452,1101479\},                                         
and an STSM grant under                                                      
COST Action IC1205.                                                          
This work was done in part while G.~Erd\'{e}lyi was visiting the             
University of Rochester.}}
\author{
G\'{a}bor Erd\'{e}lyi\\
School of Economic Disciplines\\
University of Siegen\\
57076 Siegen, Germany\\
erdelyi@wiwi.uni-siegen.de
\and
Edith Hemaspaandra\\
Department of Computer Science\\
Rochester Institute of Technology\\
Rochester, NY 14623, USA \\
www.cs.rit.edu/$\sim$eh
\and Lane A.~Hemaspaandra\\
Department of Computer Science\\
University of Rochester\\
Rochester, NY 14627, USA \\
www.cs.rochester.edu/u/lane%
}
\date{October 7, 2014}

\maketitle
\begin{abstract}
``Control'' studies attempts to set the outcome of elections through the
addition, deletion, or partition of voters or candidates.  The set of
benchmark control types was largely set in the seminal 1992 paper by
Bartholdi, Tovey, and Trick that introduced control, and there now is
a large literature studying how many of the benchmark types various
election systems are vulnerable to, i.e., have polynomial-time
attack algorithms for.

However, although the longstanding benchmark models of addition and
deletion model relatively well the real-world settings that inspire
them, 
the longstanding benchmark models of partition 
model
settings that are arguably 
quite distant from those they seek to 
capture.

In this paper, we introduce---and for some important 
cases
analyze the complexity of---new partition models that seek to better 
%
capture many real-world partition settings.  In
particular, in many partition settings one wants the two parts of the
partition to be of (almost) equal size, 
or is partitioning
into more than two parts, or has groups of actors who
must be placed in the same part of the partition.
Our hope is that having these new partition types 
will allow studies of
control
attacks to include such
models that 
more realistically 
capture many settings.




\end{abstract}

\section{Introduction, 
Motivation, and Discussion}\label{s:intro}
Elections are an important framework for decision-making, both in
human settings and in multiagent systems settings as varied as 
recommender systems~\cite{gho-her-mun-sen:c:voting-for-movies},
rank aggregation and 
web-spam filtering~\cite{dwo-kum-nao-siv:c:rank-aggregation},
similarity search and classification~\cite{fag-kum-siv:c:similarity-search},
and planning~\cite{eph-ros:j:multiagent-planning}.

Given elections' importance, it is natural that people and other 
agents should
attempt manipulative attacks on such systems, and that computational
social choice researchers should investigate the computational
complexity of conducting such attacks.  Such studies have focused 
primarily on three broad streams of manipulative attacks, known as 
manipulation, bribery, and control.  

Control, which is the focus of this paper because among the three attack
streams its
current model is by far the most troubled as to naturalness, studies
whether by changing the structure of an election---adding or deleting
or partitioning voters or candidates---an actor can ensure that a
given candidate wins.  A set of 11 benchmark attacks---``control by
adding voters,'' ``control by deleting candidates,'' etc.---are often
studied to seek to understand which types of attacks a given election
system computationally resists.  The 11 benchmark attacks 
%
already were present in the seminal 
Bartholdi, Tovey, and Trick~\shortcite{bar-tov-tri:j:control}
paper on control, at least as refined in 
subsequent papers
that made the partition tie-breaking options
explicit \cite{hem-hem-rot:j:destructive-control}
and addressed the one asymmetry in the seminal 
paper's definitions \cite{fal-hem-hem-rot:j:llull}.

With these benchmark types in hand, many election systems have been
evaluated as to how many of these types of attacks they
computationally resist, with the goal of identifying natural election
systems that resist as many as possible of the 11 benchmark types---or 
the 22 (or 21, due to a collapse of types 
recently noticed by 
Hemaspaandra, Hemaspaandra, and Menton~\shortcite{hem-hem-men:t3PlusPointerToSTACSOutByConfToAppear:search-versus-decision}) such types, if one also looks at the ``destructive'' cases
where the goal is instead to prevent a given candidate from winning.

For example, among the papers that do excellent, detailed analyses
tallying how many resistances to control 
attacks are possessed by such broadly-control-resistant election systems as 
Bucklin, fallback, ranked pairs, Schulze, 
sincere-strategy preference-based approval, and normalized
range voting
are 
\cite{
par-xia:c:ranked-pairs,men-sin:c:schulze,men:j:range-voting,erd-now-rot:j:sp-av,erd-fel-rot-sch:jtoappear:control-in-bucklin-and-fallback-voting}.

However, although the benchmark control models regarding
addition/deletion of candidates/voters are relatively natural, the
benchmark models of partitioning have always been far less so.  For
many natural, real-world settings, the longstanding benchmark partition
models simply don't come close to capturing the settings that are
routinely used to motivate them.  And so, despite the fact that an
enormous amount of effort---in most of the above papers and very many
others---has been focused on the standard partition models over the
more than two decades since they were created, we feel that it is
valuable to revisit the issue of how to best 
frame 
models of partition.
Perhaps such a revisitation should have occurred ten or fifteen years
ago, but since we don't possess a time machine, the best we can do is
to approach the issue now, and in this paper we do so.

In truth, the appropriateness of a model of partition will depend
heavily on the setting one is trying to model, and there indeed are
some settings that are well-modeled by the benchmark partition types.
Still, there are 
numerous
settings that are not well-modeled by
them, and so in this paper we present, and give some initial results
regarding, new types of partitioning that we feel are worth studying
as capturing naturally important notions of partitioning.
In particular, we define three new general flavors of partitioning.

Before discussing
the three new flavors, we 
should briefly
describe control by partition.  In the classic version of
control by partition of voters, one is given the votes of each voter
(most typically as a tie-free ordering, e.g., ``Nader$>$Kerry$>$Bush'') and a
distinguished candidate one is interested in, and one asks whether there is
a way of partitioning the voters into $V_1$ and $V_2$ such that if one
has subelections among the candidates by $V_1$ and $V_2$, and then a
final election by all the voters with the candidates being the winners
of those subelections, the distinguished candidate is the winner.  In
the classic version of candidate partitioning (known as
``runoff partition of candidates''), the input is the same, but the
question is whether there is a partition of the candidates into $C_1$
and $C_2$ such that if all the voters have subelections regarding
$C_1$ and regarding $C_2$, and then 
a final election by all 
voters over
just the winners of those subelections, the distinguished candidate is
the winner.

The three variants we suggest and study,
in order to create models that are often
closer to the real-world settings that partition seeks to capture, are
equipartition, multipartition, and partition by groups.
In equipartition,
the two parts of partitions must be of the same size (or within one if
the things being partitioned are odd in cardinality).  The motivation
for this is that in real-world settings, such as apportioning people
into districts, it is very common to want the districts to be of
essentially equal size.  Even when breaking alternatives into two
groups,
%
it is natural to keep the playing field somewhat
level by expecting the groups to be of essentially equal sizes.  Yet
the classic models of partition have no constraint at all on the sizes
of the parts of the partition; although it might in reality outrage
people were this to happen, in the classic candidate-partition model
it is completely legal to partition the candidates into ${c}$ and
$C-\{c\}$, so that candidate $c$ gets a free pass into the final
election, and it is completely legal in the classic voter-partition
model to divide a population 100,000,000 state into one district having 3
voters and one district having all the rest of the voters.  Can such a
lopsided division ever be natural?  Well, actually, yes, for example if
our partitioning regards a party-based primary and there are 3 people
in the Green party and the rest are in the Libertarian party.  
But in
many settings, size-balanced partitioning is the natural and indeed
compelling expectation.  The tricky thing regarding discussing what is
natural and what model is best to capture a given real-world setting
is that these issues are highly contextual---the richness of the real
world 
means that no single model will capture all cases.  Our
goal in this paper, thus, is not to give a new model that we claim
will capture all cases, but rather to give new models that 
in many settings are 
natural and attractive, and whose addition
as models significantly broadens the class of cases one has good models
for.

If one is dividing a state or entity up into electoral districts, it
may well be the case that having just two districts 
is simply not appropriate---perhaps the law 
sets the number 
of districts to a different number.  Thus our second new model for the 
study of partition-control is
\emph{multipartition}, that is, partitioning not into two parts but into $k$
parts.  

Our third new model is partition by groups.  In partition by groups,
each actor has a color, and all actors having the same color must be
put in the same part of the partition as each other.  For example,
regarding voter partitioning, if we are breaking an electorate into
primary districts and it is forbidden to have voters who live in the
same address/apartment building (or the same block) be in different
districts, then we would make color groups for each address/apartment
building (or block).  Or regarding candidate partitioning, if in a
department the chair is splitting hiring candidates up for a series of
culling votes and it would be openly insane for candidates from the
same subfield to be put into separate culling votes, each subarea
would have a color.  Control by groups is so natural that one might
wish to study it not just for partition cases but also for 
addition/deletion
of voters/candidates, and so we provide a result for 
adding and deleting voters by groups in this paper's section on groups.
We mention that group-based (in a somewhat different framing) 
control by addition of voters 
has been previously
introduced by Chen et al.~\cite{che-fal-nie-tal:c:combinatorial-control},
and in this paper's section on groups we discuss that interesting work and 
its relationship to the present work.

In this paper, we look at these three new models for partitioning, but
due to space limitations and for simplicity, we don't seek to prove
results about simultaneous combinations of them.  However, certainly
some real-world setting will draw on combinations and so that might in
the future be a potential area for study.

Although in this paper the new models themselves are very important,
we also, for each of them, provide one or more results as to the
complexity of control within that model, especially for the most
important election system, plurality.
In some cases, our versions of control leave the
classic control problem's complexity unchanged (although for P 
cases it often
takes far more complicated algorithms to handle the new cases while
remaining in P), in some cases our versions increase the complexity
from P to NP-complete, and we have even built a system where
our version lowers the complexity from NP-complete to $\p$.  

For example, regarding equipartition, we show that for plurality,
approval, and Condorcet voting, every existing P result can be
reestablished even for equipartition.  However, we show that for
weakCondorcet elections, equipartitioning turns the P classic case into
NP-completeness.  
We prove that control by groups often jumps P classic cases up to
NP-completeness.  
And regarding multipartition and plurality, we build
a P algorithm for the most important case.

The remainder of the paper is organized as follows.
The next section discusses whether raising, lowering, or
maintaining complexity are good or bad things.  A brief definitions
section follows that.
In the three
sections after that, we cover each of our three new models, giving
formal definitions of each as well as 
results on their behavior.  The conclusions and open problems section 
ends the paper proper, followed by an appendix.

\section{Meaning of Increasing or Lowering Complexity}\label{s:meaning}
Is it good if, for a given problem, our equipartition variant is harder
than the classic partition version of the problem?  Is it good if our
version keeps the complexity the same?  Is it good if our version
lowers the complexity?

The simple answer is that there is no simple answer, and that the
questions are themselves simplistic.  If one views complexity as
trying to ``shield'' elections from undesirable 
attacks, then having high levels
of complexity for a control problem is a good thing.  If one is an
attacker such as a campaign strategist---or if one is using an election-attack
problem to model a real-world problem that one wants to solve (as for
example the election problem ``bribery'' can be used to model
resource-allocation problems)---then having polynomial-time algorithms
is a good thing.  

So whether high levels of complexity of control are desirable or
undesirable is highly perspective-based and highly setting-based.
What computational social choice theorists can do, however, is make
clear what the control complexity levels are for the most important
control types and the most important election systems.  Having that
knowledge in hand will allow social choice theorists, election-system
choosers, campaign strategists, and others to all openly see the lay of the
land.
Although we have had colleagues tell us they are troubled that efficient
algorithms could facilitate 
what they consider unethical
election manipulation, our view is that having it openly known what
systems have what weaknesses will for example let election-system
choosers adopt systems that are not vulnerable to whatever attacks
they most fear.  We feel that, just as in cryptography, the right
approach is not a head-in-the-sand one, but one of open inquiry and
discovery, so that not just ``bad'' people (or the government) know
what is or isn't vulnerable.

As a final comment here, we mention that this paper does contain
NP-completeness results.  NP-completeness is a worst-case theory, and so
for our NP-hard cases seeking results for 
other notions of hardness would indeed be
interesting.  However, as always, despite the empirical evidence that
heuristics often do well on SAT and even on 
many NP-hard election
problems (see Rothe and Schend
\shortcite{rot-sch:j:typical-case-challenges}), it is settled
theoretical truth that (unless the polynomial hierarchy collapses) no
heuristic algorithm for any NP-hard problem can asymptotically have a
subexponential error rate (see Hemaspaandra and
Williams~\shortcite{hem-wil:j:heuristic-algorithms-correctness-frequency}
for the details on this result and a discussion of how to square it
with the 
strong empirical
performance of heuristics).
Perhaps even more to the point, the majority of the results mentioned
in this paper are not about NP-hardness but are about showing that,
even for partition-control variants that might seem likely to increase
control complexity, polynomial-time control algorithms do exist.

\section{Definitions}\label{s:defs}
An election system is a mapping that, 
given the candidates and the votes, outputs 
a subset of the candidates, who are said to be 
the winners under that election system.  
We will often use the symbol $\electionsystem$ to denote an 
election system.
For approval elections, 
voters give a 0 or 1 to each candidate, and the candidate(s) 
having the largest number of 1 votes is the winner(s).
For the other election systems that we will study, votes are tie-free 
linear orderings, e.g., ``Nader$>$Kerry$>$Bush.''  In plurality elections,
whichever candidate(s) is in the top spot on the most votes 
is the winner(s).  In Condorcet (resp.,\ weakCondorcet) elections,
each candidate who is preferred to each of other candidate $d$ in
strictly more than half (resp.,\ greater than or equal to half) 
the votes is a winner.

For conciseness, we sometimes use 
bracket notation (borrowed from linguistics) 
for \emph{independent} choices, e.g., 
``The $\bthree{ball}{book}{car}$ is $\btwo{red}{heavy}$''
is shorthand for the natural six
claims obtained by making each possible choice.

\section{Equipartition}\label{s:equipartition}
Let us now define 
our equipartition notion and the classic partition problems.  
(In all our problem definitions, 
we include ``represented via preference lists over $C$.'' But 
in fact the voters are always represented by whatever vote type
is that of the election system; throughout this paper that 
is always preference lists, 
except for approval voting the votes are instead 0/1-vectors.)
Recall that partitioning a set into two (or more)
parts means that every element of the set must appear in exactly one of 
the parts.

\EP{$\cale$-CCREPC (Control by Runoff Equipartition of Candidates)}
{A set $C$ of candidates, 
a 
collection
$V$ of voters
  represented via preference lists over $C$, and a distinguished candidate
  $p \in C$.}
{Is there
  a partition of $C$ into $C_1$ and $C_2$ such that 
  $| \, \|C_1\| - \|C_2\| \, | \leq 1$ and 
$p$ is the sole winner of
  the two-stage election where the winners of subelection $(C_1,V)$
  that survive the tie-handling rule compete against the winners of
  subelection $(C_2,V)$ that survive the tie-handling rule? Each
  subelection (in both stages) is conducted using election
  system~$\electionsystem$.}

\EP{$\cale$-CCEPV (Control by Equipartition of Voters)}
{A set $C$ of candidates, a 
collection
$V$ of voters
  represented via preference lists over $C$, and a distinguished candidate
  $p \in C$.}
{Is there
  a partition of $V$ into $V_1$ and $V_2$ such that 
  $| \, \|V_1\| - \|V_2\| \, | \leq 1$ and 
$p$ is the sole winner of
  the two-stage election where the winners of election $(C,V_1)$ that
  survive the tie-handling rule compete against the winners of
  $(C,V_2)$ that survive the tie-handling rule?  Each subelection (in
  both stages) is conducted using election system~$\electionsystem$.}

The classic cases, $\cale$-CCRPC
and 
$\cale$-CCPV, are defined
identically, except without the clause forcing the partition parts to
be equal in size (or off-by-one if the set being partitioned is of odd
cardinality).

For all of the above, there is the
issue of whether if there are multiple winners of a subelection, all
of them move on to the final election (called ties-promote, notated TP)
or none of them move on to the final election (called ties-eliminate, notated
TE; in this model, to move on to the final election one must be the
unique winner of a subelection).  Thus each problem will always 
appear with a TE or TP to specify the tie-handling approach, e.g.,
$\cale$-CCPV-TE or 
$\cale$-CCPV-TP\@.  

The literature also contains a ``bye'' version of 
partitioning candidates, in which 
any number of candidates can be assigned to skip (``bye'') 
the first round and the rest compete to get into the second round.  
Since equipartition is not natural for that ``bye''
version (because the number of candidates skipping a first round is usually 
driven by such things as 
excesses relative to powers of two in the number of candidates),
we have not defined here either the classic or an ``equi'' version of 
``bye'' partition;  
a later version of this paper, however, will note that 
many of our results hold for the ``equi'' version of ``bye'' partition,
and more importantly will give, and prove results about, a model
in which one specifies as part of the input how 
many candidates get a ``bye,'' since that model provides
the natural partition-size-sensitive variant of the ``bye'' 
candidate-partition 
problem.

As a final comment about model details, all of the partition 
problems discussed above 
are in what is called the unique-winner model, i.e., the goal is 
to make a given candidate the one and only overall winner.  That is the 
model of the seminal Bartholdi, Tovey, and 
Trick~\shortcite{bar-tov-tri:j:control}
control paper and all the immediately subsequent control papers, 
and probably even today is the most common model when studying 
control (although we ourselves prefer the alternate model 
known as the nonunique-winner model or the co-winner model, 
where one merely needs to 
make a given candidate
become \emph{a} winner).  
And so throughout all parts of this 
version, 
we focus solely on the 
unique-winner model, since
doing so creates 
apples-to-apples 
contrasts for those cases where our models
change the existing complexity from those found in 
that seminal paper and various other papers.
(We have checked the vast majority of our results in both models, and in 
a later version of this paper will cover both models.  This is not 
an issue one can safely take 
for granted, since there are examples in the literature where the 
complexity or behavior of the two models differs 
sharply, e.g., \cite{fal-hem-sch:c:copeland-ties-matter,hem-hem-men:t3PlusPointerToSTACSOutByConfToAppear:search-versus-decision}.)  


Let us turn to our results, which give a sense of what holds when one
partitions while required to have the parts be ``equi.''  We show that
for many important systems, including approval, Condorcet, and plurality, 
partition problems remain in P even for equipartitioning,
albeit typically with substantially more difficult algorithms and new
tricks relative to what the non-``equi'' cases required.  However, in
contrast to its sibling, the Condorcet case, 
we prove that for weakCondorcet an
increase from P to NP-completeness occurs.  We also construct 
an (admittedly artificial) 
election system having its winner problem in P, such that 
going from partition to equipartition lowers the control complexity from
NP-completeness to $\p$.

The following shows that 
for 
approval, Condorcet, and plurality,
each P case
of 
CC$\btwo{RPC}{PV}$-$\btwo{TE}{TP}$---these can be found in Table~1 of 
Hemaspaandra, Hemaspaandra and Rothe~\shortcite{hem-hem-rot:j:destructive-control} and are 
variously due to that paper and Bartholdi, Tovey,
and Trick~\shortcite{bar-tov-tri:j:control}---remains in P for equipartition.

\begin{theorem}\label{t:equi-down}
Each of the problems \textrm{\rm{}Plurality-CCEPV-TE} and 
\textrm{\rm{}$\btwo{Condorcet}{approval}$-CCREPC-$\btwo{TE}{TP}$} 
belongs to $\p$.
\end{theorem}

In the appendix, we include a proof of
$\textrm{\rm{}Plurality-CCEPV-PE} \in \p$.  The proof of this has two
interesting issues that do not occur in the standard partition case.

First, even for those inputs where $p$ is at the top of no more than
$\lceil \|V\|/2\rceil$ of the votes, one cannot assume within the
proof that if candidate $p$ can be made to win, it can be made to win
with some partition that puts all votes with $p$ at their top in the
same partition part.  Here is an example showing that that assumption,
which
works in the general case, fails here.
If we
 have 5 votes for $p$, $6$ votes for $a$, and 3 votes for $b$, we can
 make $p$ a sole overall winner by letting $V_1$ consist of 4 votes for $p$ and 3
 votes for $a$.  Then $p$ is the unique winner in $V_1$, and $V_2$ consists of 
 3 votes for $a$, 3 votes for $b$, and 1 vote for $p$, so no candidate 
 from the second election makes it through to the runoff, and $p$ is 
 the overall unique winner.  However, if we put all 5 votes for $p$ in 
 $V_1$, then since there are 7 votes in $V_1$, at least 4 votes for $a$ 
 are in $V_2$ and $a$ is the unique winner of the second subelection, and
 so also of the overall election. 

The second twist is that we need to find a ``safe'' way to legally
distribute in an ``equi'' overall fashion certain 
``remaining'' votes; and
our proof does this by pushing them all to one side (violating
``equi,'' typically), and then correcting this in a way that is
guaranteed to succeed if success is possible.

We have also shown that the NP-complete cases still hold 
for our systems of interest.
\begin{theorem}\label{t:np-stays-np}
For 
approval, Condorcet, and plurality,
each $\np$-complete case
of 
{\rm{}CC}$\btwo{RPC}{PV}$-$\btwo{TE}{TP}$---these can be found in Table~1 of 
Hemaspaandra, Hemaspaandra and Rothe~\shortcite{hem-hem-rot:j:destructive-control} and are 
variously due to that paper and Bartholdi, Tovey,
and Trick~\shortcite{bar-tov-tri:j:control}---remains $\np$-complete
for the one among 
{\rm{}CC}$\btwo{REPC}{EPV}$-$\btwo{TE}{TP}$
that is its equipartition analogue.
\end{theorem}

Now, it might be natural to wonder: Is there a meta-theorem showing
that NP-completeness always inherits from partition cases to 
equipartition cases?  If so Theorem~\ref{t:np-stays-np} would become 
a freebie consequence of the meta-theorem.  However, although we do 
not have any example of a natural election system where the 
``equi'' case drops the complexity from NP-completeness to P, 
we have constructed an election system displaying precisely that 
behavior.  
And so NP-completeness does not 
always 
inherit from the
standard case to the ``equi'' case. Theorem~\ref{t:nontransfer}'s
proof is included in the appendix.

\begin{theorem}\label{t:nontransfer}
There exists an election system $\electionsystem$, whose winner 
problem is in $\p$, such that {\rm{}$\cale$-CCPV-TP} is $\np$-complete,
yet 
{\rm{}$\cale$-CCEPV-TP} belongs to $\p$.
\end{theorem}

In contrast to its close relative, Condorcet elections, weakCondorcet
elections increase complexity from the 
partition case to the equipartition case 
for the RPC-TP case.  
(This complexity increase 
is not precluded
by the general fact that 
subcases of problems cannot be harder than the original problem.
Although each equipartition of a set is indeed a partition of that set,
we are not dealing here with 
a subcase of a problem, but rather with a control problem whose allowed 
internal actions are a subset of those of a different control problem,
and so there is no automatic prohibition on the complexity increasing.)
We do not yet have a complexity
classification for 
weakCondorcet-CCREPC-TE, and consider
that an interesting open problem.

\begin{theorem}\label{t:wc}
$\textrm{\rm{}weakCondorcet-CCRPC-}\btwo{TE}{TP}$ are in $\p$, 
but 
$\textrm{\rm{}weakCondorcet-CCREPC-TP}$ is $\np$-complete.
\end{theorem}

Briefly, what is behind the change in complexity 
here is 
%
that we can make one of the subelections represent a vertex cover 
and we use equipartition to limit the size of that vertex cover.
The reason the same approach does not work also for Condorcet 
elections is that we crucially need 
that we can have multiple winners.
We now prove the two TP parts of Theorem~\ref{t:wc}, as these 
are what bring out the contrasting behavior.



\begin{proofs}
To show that weakCondorcet-CCRPC-TP is in $\p$,
it suffices to 
note that for all $p \in C$, if
$p$ can be made the unique weakCondorcet winner in $(C,V)$ by RPC-TP,
then this is established by candidate partition $(\{p\}, C-\{p\})$.  
To see this, suppose $p$ is not the unique weakCondorcet winner in
partition $(\{p\}, C-\{p\})$.   Then there is a candidate $c \neq p$ such
that $c$ is a weakCondorcet winner in $(C-\{p\},V)$ and $c$ ties-or-defeats
$p$ in their head-to-head contest.  But then $c$ is a weakCondorcet winner
in $(C,V)$ and so in any $(C',V)$ with $C' \subseteq C$ and $c \in C'$.
It follows that $c$ is always a winner by RPC-TP, and so $p$ will never be the
unique winner.

That was easy.  But $(\{p\}, C-\{p\})$ is clearly not an equipartition.
We will now show that for 
weakCondorcet-CCREPC-TP is NP-complete. 
We will prove this by a reduction from Cubic Vertex Cover:
Given a graph $G = (V,E)$ that is cubic, i.e., where every vertex has degree
three, and a positive integer $k \leq \|V\|$,
we ask whether $G$ has a vertex cover of size $k$, i.e.,
a set of vertices $V' \subseteq V$ of size $k$
such that every edge in $E$ is incident with at least one vertex in $V'$.
Let $\|V\| = n$. Since $G$ is cubic, $\|E\| = 2n/3$.

Using McGarvey's construction~\cite{mcg:j:election-graph}, we construct, in
polynomial time, an election $(C,\widehat{V})$ with
the following properties:
\begin{itemize}
\item $C = \{p\} \cup V \cup E \cup D$, where
$D = \{d_1, \ldots, d_{n/2 + 2k - 1}\}$.
\item The set of voters, $\widehat{V}$,
is such that we have the following head-to-head 
contest results:
\begin{itemize}
\item for every $e \in E$, $e$ defeats $p$,
\item for every $c \in V \cup D$, $p$ defeats $c$,
\item for every $e = \{v,v'\} \in E$, 
$v$ defeats $e$ and $v'$ defeats $e$, 
\item all other head-to-head contests are ties.
\end{itemize}
\end{itemize}

Suppose $V'$ is a vertex cover of size $k$ of $G$.  Then $p$ can be made the
unique weakCondorcet winner by REPC, using partition
$(\{p\} \cup D \cup V-V', E \cup V')$.  Note that
$\|\{p\} \cup D \cup V-V'\| = 1 + n/2 + 2k - 1 + n-k =
3n/2 + k = \|E \cup V'\|$. $p$ is the Condorcet winner in
$(\{p\} \cup D \cup V-V', \widehat{V})$, and thus participates in the
runoff.  
Since $V'$ is a vertex cover, for every candidate $e \in E$, there
is a candidate $v \in V'$ such that $v$ defeats $e$ in their head-to-head
contest.  So no candidate in $E$ makes it to the runoff.
So $p$ is the Condorcet winner in the runoff, and thus
certainly the unique weakCondorcet winner.

For the converse, suppose $p$ can be made the unique weakCondorcet winner
by REPC-TP\@.  Let $(C_1, C_2)$ be an equipartition of $C$ with
$p \in C_1$ that witnesses this.  Then $p$ is a weakCondorcet winner
in $(C_1,\widehat{V})$.  This implies that $E \subseteq C_2$.  Since
$p$ is a weakCondorcet winner in the runoff, no candidate from $E$ participates
in the runoff.  So for every $e \in E$, there is a $c \in C_2$ such that
$c$ defeats $e$ in their head-to-head contest.
The only candidates that defeat $e = \{v,v'\}$ are
$v$ and $v'$.  It follows that $C_2 \cap V$ is a vertex cover
of $G$.  Since $(C_1,C_2)$ is an equipartition, 
$\|C_2 \cap V\| \leq k$.
\end{proofs}

\section{Multipartition}\label{s:multipartition}
In many settings two simply is not the number of parts into which
one's voter set must be divided.  For example, the Dean may wish to
have three study sections each passing forward a best choice on some
issue.  Multipartition, which we'll define here just for PV but it can
just as well be defined for RPC, generalizes the 2-partition PV
problem used in the seminal Bartholdi, Tovey, and Trick paper to each
$k$-partitioning.  For each integer $k \geq 2$, define the following 
problem.

\EP{$\cale$-CCP${}_k$V (Control by $k$-partition of Voters)}
{A set $C$ of candidates, a 
collection
$V$ of voters
  represented via preference lists over $C$, and a distinguished candidate
  $p \in C$.}
{Is there
  a partition of $V$ into $k$ parts, $V_1,V_2,\ldots,V_k$, such that 
$p$ is the sole winner of
  the two-stage election where the winners of each of the $k$ 
  elections $(C,V_i)$ 
  that
  survive the tie-handling rule compete against each other
  in a final election?  Each subelection (in
  both stages) is conducted using election system~$\electionsystem$.}

Plurality-CCP$_2$V-TE is in P \cite{hem-hem-rot:j:destructive-control}.
As Theorem~\ref{t:multi}
states, we in fact have that P membership still holds for each 
$k$-partition version.
The proof is by breaking things into a huge but polynomial
number of cases by for each part of the partition either 
guessing a candidate who allegedly uniquely wins that part along with the 
score that candidate has within that part
or guessing a pair of candidates who 
tie (perhaps along with others) as the winners of that part of the 
partition along with the score those two achieve in that part, 
and then for each such
case appropriately checking in polynomial time whether it can be realized.

\begin{theorem}\label{t:multi}
For each $k \geq 2$, 
$\textrm{\rm{}Plurality-CCP$_k$V-TE}
\in \p$.
\end{theorem}

It would be interesting to study multipartition for other 
election systems, and also to study multipartition varied to 
allow the number of partitions to itself not be fixed 
but rather to be specified as part 
of the input.

\section{Voter Control by Groups}\label{s:groups}
In voter partition by groups, each vote has a color (i.e., a label),
and all votes with the same label must be placed into the same
partition part.  We also define group voter-control problems for deleting
voters (where all votes of a given color must be jointly deleted or
kept) and for adding voters (where in the pool of potential additional
voters each one has a color, and each color group must be added or not
added as a block).  As discussed in the introduction, these models
capture cases where groups cannot be separated.  One example
might be due to living at
the same address in a redistricting problem, and another might be 
a departmental study group process in which each of the department's 
area subfaculties must be placed within the same study group.
We below define just the 
voter cases for control by groups, but one could 
completely analogously define candidate control by groups.

\EP{$\cale$-CCPVG (Control by Partition of Voter Groups)}
{A set $C$ of candidates, a collection $V$ of voters represented
via  preference lists over $C$, 
a partition of $V$ into any number of groups $G_1,\ldots, G_k$,
and a distinguished candidate $p\in C$.}
{Is there
  a partition of $V$ into $V_1$ and $V_2$ such that 
for each $i$ either $G_i \subseteq V_1$ or $G_i \subseteq V_2$ holds
and
$p$ is the sole winner of
  the two-stage election where the winners of election $(C,V_1)$ that
  survive the tie-handling rule compete against the winners of
  $(C,V_2)$ that survive the tie-handling rule?  Each subelection (in
  both stages) is conducted using election system~$\electionsystem$.}

\EP{$\cale$-CCDVG (Control by Deleting Voter Groups)}
{A set $C$  of candidates, 
a 
collection
$V$ of voters
  represented via preference lists over $C$, 
a partition of $V$ into any number of groups $G_1,\ldots, G_k$,
a nonnegative integer $\ell$, 
and a distinguished candidate
  $p \in C$.}
{Is there a set $S \subseteq V$, $\|S\| \leq \ell$,
such that 
$p$ is the sole winner of the $\cale$ election over $C$ with the vote 
collection set being $V$ with $S$ (multiset) removed,
and 
for each $i$ either $G_i \subseteq S$ or $G_i \cap S = \emptyset$?}

\EP{$\cale$-CCAVG (Control by Adding Voter Groups)}
{A set $C$  of candidates, 
a 
collection
$V$ of voters
  represented via preference lists over $C$, 
a 
collection
$W$ of potential additional voters
  represented via preference lists over $C$, 
a partition of $W$ into any number of groups $G_1,\ldots, G_k$,
a nonnegative integer $\ell$, 
and a distinguished candidate
  $p \in C$.}
{Is there a collection $S \subseteq W$, $\|S\| \leq \ell$, 
such that 
$p$ is the sole winner of the $\cale$ election over $C$ with the vote 
set being $V$ (multiset) unioned with 
$S\!$, and 
for each $i$ either $G_i \subseteq S$ or $G_i \cap S = \emptyset$?}

%
Before stating results for this model, let us quickly discuss
whether these notions are in overlap with models in the literature.
After all, votes are coming and going in blocks, and so one might
wonder if this is related to for example the notion of 
weighted control introduced by 
Faliszewski, Hemaspaandra, and Hemaspaandra
\shortcite{fal-hem-hem:c:weighted-control}.  Briefly, 
the notions are different to their core in that a weighted vote 
puts a lot of weight \emph{on that vote}, but in contrast, a 
vote group may consist of votes that have vastly different 
preferences from each other.  It is true that if one took the 
Faliszewski, Hemaspaandra, and Hemaspaandra
\shortcite{fal-hem-hem:c:weighted-control} notion of weighted 
control, and restricted the weights to being input in unary,
and for the adding/deleting voters cases shifted from that 
paper's model of counting as one's limit the number votes and instead 
adopted the model (mentioned but not adopted in that paper) of limiting 
by the total weight of votes added/deleted, then \emph{those} weighted 
control problems would each indeed many-one polynomial-time reduce to
our analogous voter control by groups problem;  but that seems to be 
far as the connection goes between the two papers.

A closer connection is to the work of Chen et
al.~\cite{che-fal-nie-tal:c:combinatorial-control}, who define and
study a very general notion of ``combinatorial voter control'' for
addition of voters.  (Their paper is not concerned with partition 
problems, the main focus of the present paper.)
Loosely put, for each voter they have a group of
voters who in some sense follow that voter, so that if one takes an
action on a voter, the group of the voter follows also.  Note that
this is a very flexible, general scheme, and for example does not
require that the follower function
breaks the voters into equivalence classes, as
does our coloring scheme.  Of course, when proving NP-hardness results,
such flexibility \emph{weakens} the results.  So in their paper 
(which is in the nonunique-winner model) they define and
study a number of restricted models of the follower function.
However, even the most restrictive models of follower functions that they 
study are incomparable to our model (even when they add 
nice symmetry-like properties, they focus those on voters with 
the same preferences, and so are not focused on what we are focused on,
which is, in effect, coloring voters, i.e., breaking
voters into equivalence classes in whatever arbitrary way is specified 
by the coloring), and so our NP-hardness result for 
Plurality-CCAVG is incomparable with their NP-hardness results for 
their model of combinatorial control by adding voters.  Nonetheless,
it is important to mention 
their work appeared in conference two months before any version 
of our paper appeared, and so their paper certainly deserves the 
precedence and credit for the idea of grouping voters, and we mention
that their paper establishes very interesting results---both 
hardness and easiness---for voter-addition control 
for a broad range of types of follower functions (although not 
for the case we study here).

Turning to our results for our model, unlike our earlier two models,
the addition of groups very broadly increases P complexity levels to
NP-completeness.  For the three cases covered by the following 
theorem, the analogous result without groups is well known to 
be in P~\cite{bar-tov-tri:j:control,hem-hem-rot:j:destructive-control}.
And the NP-completeness claims of Theorem~\ref{t:groups}
are each proved by building an appropriate reduction from
an NP-complete problem, in particular Exact Cover by 3-Sets.
\textrm{\rm{}Plurality-CCPVG-TP} is also NP-complete, but we do not 
state that in the theorem since this follows 
immediately
from the known
result (see Hemaspaandra, Hemaspaandra, and 
Rothe \shortcite{hem-hem-rot:j:destructive-control}) that 
\textrm{\rm{}Plurality-CCPV-TP} is NP-complete.

\begin{theorem}\label{t:groups}
Each of \textrm{\rm{}Plurality-CC$\bthree{PVG-TE}{AVG}{DVG}$}
is $\np$-complete.
\end{theorem}
In the appendix, we include a proof of
the PVG-TE case.

\section{Conclusions and Open Directions}
We introduced three models of partition control---equipartition,
multipartition, and partition by groups---that seek to for many cases
more closely model real-world situations than the twenty-year-old
standard benchmark set.  We obtained a number of results 
on the
complexity of our new models with respect to important election
systems, especially plurality, the most prevalent of election systems.
We established many natural examples where the variants are of the
same complexity as their analogous standard benchmark model, and also
established many natural examples where the variants' complexity
increases relative to the analogous standard benchmark model.

The current version of the paper focused on the unique-winner model and
so-called constructive control, but 
a later version 
will 
cover the nonunique-winner model (in which our
results broadly still hold) and 
so-called destructive control.

Open directions include studying combinations of 
our new models, seeking additional models to better capture 
real-world settings, doing studies 
to determine how well various models do capture
real-world settings, 
extending the present study to partial-information 
models, seeking typical-case hardness results, 
pursuing the additional multipartition studies mentioned
at the end of the multipartition section,
and resolving the
complexity of weakCondorcet-CCREPC-TE to see whether it provides 
an additional natural example of an increasing complexity level (see
Theorem~\ref{t:wc} and the comments preceding it).



\clearpage



\appendix

\section{Appendix:
Selected Additional Proofs}


\subsection{Proof of the Plurality Part of Theorem~\ref{t:equi-down}}
We now prove the claim that Plurality-CCEPV-TE belongs to $\p$.

\begin{proofs}  

The example given in the main text immediately after the 
statement of Theorem~\ref{t:equi-down} shows that 
we have to be really careful about what assumptions we 
make in our proof.  However, we
indeed can show that our control problem is in $\p$.  First of all, note that
$p$ can be made a unique winner by EPV-TE if and only if there exists an
equipartition $(V_1,V_2)$ such that $p$ is the unique winner of 
$(C,V_1)$ and 
\begin{enumerate}
\item $(C,V_2)$ has $c$ as unique winner and $p$ defeats $c$, or
\item $(C,V_2)$ has $p$ as a winner, or
\item $(C,V_2)$ has more than one winner.
\end{enumerate}

These three conditions can be checked as follows in polynomial time.
\begin{enumerate}
\item
For every $c \in C -\{p\}$ such that $p$ defeats $c$,
for every $k_p \leq \score_V(p)$ and
for every $k_c \leq \score_V(c)$,
put $k_p$ votes for $p$ in $V_1$ and the remaining votes for $p$ in $V_2$ and
put $k_c$ votes for $c$ in $V_2$ and the remaining votes for $c$ in $V_1$.

We will now check whether $(V_1,V_2)$ can be extended to a desired equipartition.
If $p$ is not the unique winner in $V_1$ or $c$ is not the unique winner in 
$V_2$, then this is not possible and we move on to the next loop iteration.

Otherwise, for each $d \in C - \{p,c\}$, put as many votes for $d$ as
possible into $V_1$ while keeping $p$ the unique winner in $V_1$
(i.e., $\min(score_V(d), k_p - 1)$ votes).  

If $\|V_1\| < \lfloor \|V\|/2 \rfloor$, there are not enough 
votes in $V_1$ and we move on to the next loop iteration.
Otherwise, move votes for candidates in $C - \{p,c\}$ from $V_1$ to
$V_2$ if this is possible while keeping $c$ the unique winner in $V_2$.
Keep doing this until $(V_1,V_2)$ becomes an equipartition, in which case
we have found a successful equipartition, or until it is not possible
to move votes for candidates in $C - \{p,c\}$ from $V_1$ to $V_2$ while keeping 
$c$ the unique winner in $V_2$, in which case we move on to the next loop
iteration.
\item
This is similar to the previous case.
For every $0 < k_p \leq \score_V(p)$ 
put $k_p$ votes for $p$ in $V_1$ and the remaining votes for $p$ in $V_2$.

We now will check whether $(V_1,V_2)$ can be extended to a desired equipartition.
For each $d \in C - \{p\}$, put as many votes for $d$ as
possible into $V_1$ while keeping $p$ the unique winner in $V_1$
(i.e., $\min(score_V(d), k_p - 1)$ votes).  

If $\|V_1\| < \lfloor \|V\|/2 \rfloor$, there are not enough 
votes in $V_1$ and we move on to the next loop iteration.
Otherwise, move votes for candidates in $C - \{p\}$ from $V_1$ to
$V_2$ if this is possible while keeping $p$ a winner in $V_2$.
Keep doing this until $(V_1,V_2)$ becomes an equipartition, in which case
we have found a successful equipartition, or until it is not possible
to move votes for candidates in $C - \{p\}$ from $V_1$ to $V_2$ while keeping 
$p$ a winner in $V_2$, in which case we move on to the next loop iteration.

\item
The case that $p$ is one of the winners of $(C,V_2)$ has been handled in
the previous case, so it suffices to handle the case where $(C,V_2)$ has
at least two winners in $C - \{p\}$.

For every $c,c' \in C$ such that $\|\{p,c,c'\}\| = 3$,
for every $k_p \leq \score_V(p)$, and
for every $k_c \leq \min(\score_V(c), score_V(c'))$,
put $k_p$ votes for $p$ in $V_1$ and the remaining votes for $p$ in $V_2$,
put $k_c$ votes for $c$ in $V_2$ and the remaining votes for $c$ in $V_1$,
and
put $k_c$ votes for $c'$ in $V_2$ and the remaining votes for $c'$ in $V_1$.

We now will check whether $(V_1,V_2)$ can be extended to a desired equipartition.
If $p$ is not the unique winner in $V_1$ or $c$ and $c'$ are not
winners in $V_2$, then this is not possible
and we move on to the next loop iteration.

Otherwise, for each $d \in C - \{p,c,c'\}$, put as many votes for $d$ as
possible into $V_1$ while keeping $p$ the unique winner in $V_1$
(i.e., $\min(score_V(d), k_p - 1)$ votes).  

If $\|V_1\| < \lfloor \|V\|/2 \rfloor$, there are not enough 
votes in $V_1$ and we move on to the next loop iteration.
Otherwise, move votes for candidates in $C - \{p,c,c'\}$ from $V_1$ to
$V_2$ if this is possible while keeping $c$ and $c'$ winners in $V_2$.
Keep doing this until $(V_1,V_2)$ becomes an equipartition, in which case
we have found a successful equipartition, or until it is not possible
to move votes for candidates in $C - \{p,c,c'\}$ from
$V_1$ to $V_2$ while keeping 
$c$ and $c'$ winners in $V_2$, in which case we move on to the next loop
iteration.
\end{enumerate}
\end{proofs}

\subsection{Proof of Theorem~\ref{t:nontransfer}}
We prove Theorem~\ref{t:nontransfer}, namely, that
there exists an election system $\electionsystem$, whose winner 
problem is in $\p$, such that {\rm{}$\cale$-CCPV-TP} is $\np$-complete,
yet 
{\rm{}$\cale$-CCEPV-TP} belongs to~$\p$.


\begin{proofs}
We define $\cale$ as follows.
On input $(C,V)$:
\begin{itemize}
\item If $\|C\| \leq 4$ and ($C \cap \{0,1,2,3\} = \{0,2\}$ or
$C \cap \{0,1,2,3\} = \{1,3\})$,
then the winners are the approval winners of $(C-\{0,1,2,3\},V)$.
\item If $\|C\| \leq 4$ and $C \cap \{0,1,2,3\} \neq \{0,2\}$
and $C \cap \{0,1,2,3\} \neq \{1,3\}$,
there are no winners.
\item If $\|C\| > 4$ and $\{0,1,2,3\} \subseteq C$,
then $\|V\| \bmod 4$ is a winner and if
$(C-\{0,1,2,3\}, V)$ has a unique approval winner, then that
candidate is also a winner.
There are no other winners.
\item If $\|C\| > 4$ and $\{0,1,2,3\} \not \subseteq C$, there
are no winners.
\end{itemize}

We first show that $\cale$-CCEPV-TP is in $\p$. This is easy.  If $\|C\|
\leq 4$, 
there are no winners in the runoff, since 0, 1, 2, and 3 do not
participate in the runoff.  If $\|C\| > 4$ and $\{0,1,2,3\} \not \subseteq C$,
there are no candidates in the runoff.  
If $\|C\| > 4$ and $\{0,1,2,3\} \subseteq C$,
there are at most four candidates in the runoff.
The candidates in $\{0,1,2,3\}$ that
participate in the runoff are exactly
$\|V_1\| \bmod 4$ and $\|V_2\| \bmod 4$ for partition
$(V_1,V_2)$.
But if $(V_1,V_2)$ is an equipartition,
it is never the case that
$\{\|V_1\| \bmod 4, \|V_2\| \bmod 4\} = \{0,2\}$
or
$\{\|V_1\| \bmod 4, \|V_2\| \bmod 4\} = \{1,3\}$
and so there are no winners in the runoff.

To show that $\cale$-CCPV-TP is NP-complete, we reduce from
approval-CCPV-TE, which is
NP-complete~\cite{hem-hem-rot:j:destructive-control}.
Let
$(C,V)$ be an election and let $p$ be the preferred candidate.
Assume that $C \cap \{0,1,2,3\} = \emptyset$.  
Let $\widehat{C} = C \cup \{0,1,2,3\}$ and let $\widehat{V}$ consist
of the voters in $V$ (extended to $\widehat{C}$ by
not approving of candidates in $\{0,1,2,3\}$)
plus two additional voters that don't approve of any candidate
if $\|V\|$ is even
and one additional voter that doesn't approve of any candidate if
$\|V\|$ is odd.
We claim that $p$ can be made the unique approval winner
in $(C,V)$ by PV-TE if and
only if $p$ can be made the unique $\cale$ winner
in $(\widehat{C},\widehat{V})$ by PV-TP\@.

First suppose that $(V_1,V_2)$ is a partition of $V$ 
that makes $p$ the unique approval winner by PV-TE\@.
If $\|V\|$ is odd, add the one additional voter that doesn't approve of any 
candidate to $V_1$ or $V_2$ in such a way that $\|V_1\| \bmod 4 
\neq \|V_2\| \bmod 4$.  
If $\|V\|$ is even and $\|V_1\| \bmod 4 = \|V_2\| \bmod 4$, add the
two additional voters to $V_1$.  
If $\|V\|$ is even and $\|V_1\| \bmod 4 \neq \|V_2\| \bmod 4$, add one
additional voter to $V_1$ and one additional voter to $V_2$.
In all cases, we now have a partition 
$(\widehat{V_1},\widehat{V_2})$ of $\widehat{V}$ with the same unique
approval winners as before
(when restricting the candidates to $C$)
and such that 
$\{\|V_1\| \bmod 4, \|V_2\| \bmod 4\} = \{0,2\}$ or
$\{\|V_1\| \bmod 4, \|V_2\| \bmod 4\} = \{1,3\}$.
This immediately implies that this partition makes $p$ the
unique winner in $\cale$-CCPV-TP\@.

For the converse, suppose  that
$(\widehat{V_1},\widehat{V_2})$ is a partition of
$\widehat{V}$ that makes $p$ the unique $\cale$ winner by
PV-TP\@. 
Then $(\widehat{V_1},\widehat{V_2})$ 
makes $p$ the unique approval
winner in $(C,\widehat{V})$ by PV-TE\@.  Now simply delete 
the additional voters that don't approve of any candidate
to obtain partition $(V_1,V_2)$ of $V$
that makes $p$ the unique approval winner in 
$(C,V)$ by PV-TE\@. 
\end{proofs}

\subsection{Proof of the PVG-TE Part of Theorem~\ref{t:groups}}
We now prove the claim that Plurality-CCPVG-TE is 
$\np$-complete.

\begin{proofs}
  We reduce from \textsc{X3C}.  Given a set $B=\{ b_1,
  \ldots,b_{3m}\}$, $m>1$, and a collection $\mathcal{S}=\{S_1,
  \ldots,S_n\}$ of subsets $S_i =\{ b_{i,1}, b_{i,2}, b_{i,3}\}
  \subseteq B$ with $\| S_i \| =3$ for each $i$, $1\leq i \leq
  n$.  

  We assume $n>m+1$. We may safely make this assumption,
  as 
    \textsc{X3C} still remains $\np$-complete under this
    restriction. $n<m$ is automatically a no instance and the two cases
    $n=m$ and $n=m+1$ are solvable in polynomial time. Thus the
    problem is still $\np$-complete under the restriction $n>m+1$.
  
  Define the election $(C,V)$, where $C=\{ p,c,d,e\} \cup B$ is the
  set of candidates, $p$ is the distinguished candidate, and $V$
  consists of the following $n+3$ groups of voters.
  As a shorthand, when specifying votes we will sometimes include 
  a set of candidates, when resolving those as any linear ordering 
  among those voters (e.g., lexicographic) will be fine for the vote's
  role in the proof.  For example, $p > S > b$, where $S = \{z,a,w\}$, 
  may be taken as a 
  shorthand for $p > a > w > z > b$.

\begin{itemize}
 \item For each $i$, $1 \leq i \leq n$, there is a group, $G_i$, 
with six voters of the form: 
\begin{itemize}
 \item $p > C-\{ p \}$, 
 \item $p > C-\{ p \}$,
 \item $b_{i,1} > C-\{ b_{i,1}, p \} > p$, 
 \item $b_{i,2}> C-\{ b_{i,2}, p \} >p$, 
 \item $b_{i,3} >C-\{ b_{i,3}, p \}>p$, and
 \item $e >C-\{ e,p\}>p$.
\end{itemize}
 \item There is a group $G_B$ consisting of the following voters:
\begin{itemize}
 \item  Let $\ell _j=\| \{S_i \in \mathcal{S} | b_j \in S_i \}\|$ 
for all $j$, $1\leq j \leq 3m$. For each $j$, $1 \leq j \leq 3m$, 
there are $2n-\ell _j$ voters of the form $b_j>C-\{ b_j,p \}>p$. 
 \item There are $2m$ voters of the form $p>C-\{ p\}$.
 \item There are $n+m-1$ voters of the form $e>C-\{ e,p\}>p$.
 \item There is one voter $v_c$ of the form $c>C-\{ c,p\}>p$. 
\end{itemize}

 \item There is a group $G_c$ containing $2(n+m)+1$ 
voters of the form $c >C-\{ c,p\} >p$.

 \item There is a group $G_d$ containing 
$2(n+m)+1$ voters of the form $d >C- \{ d,p \}>p$. 

\end{itemize}

It is easy to see that in this election each $b_j \in B$ has a
score of $2n$, candidate $c$ has a score of $2(n+m)+2$, candidate $d$
has a score of $2(n+m)+1$, candidate $e$ has a score of $2n+m-1$, and
the distinguished candidate $p$ has a score of $2(n+m)$.

We claim that $B$ has an exact cover $B'$ if and only if $p$ can be
made the unique winner of the election by control by partition of voters in
the TE model.

Suppose $B$ has an exact cover $B'$. Partition the set of voters as
follows. Let $V_2$ contain the $m$ groups corresponding to $B'$ and
the groups $G_c$ and $G_d$. Let $V_1=V-V_2$. Candidate $p$ is the
unique winner of subelection $(C,V_1)$, since 
$p$ has a score of $2n$,
each $b_j\in B$ has a score of $2n-1$, candidate $c$ has a score of
$1$, candidate $d$ has no points at all, and candidate $e$ has $2n-1$
points. There is no unique winner in subelection $(C,V_2)$, since
candidates $c$ and $d$ tie for first place, eliminating each other.
Thus only candidate $p$ moves to the final round of
the election, and $p$ is the unique winner of the final round.

For the converse direction, 
suppose $p$ can be made a winner of the election by
partition of voters in the TE model. Without loss of generality, 
assume that $p$ is the
unique winner of subelection $(C,V_1)$.  Since both candidates $c$ and
$d$ 
accumulate $2(n+m)+1$ points in their respective groups, and all the
other candidates $p$, $e$, and each $b_j \in B$, $1\leq j \leq 3m$,
have less than $2(n+m)+1$ overall points, $c$ and $d$ have to
eliminate each other in subelection $(C,V_2)$. Candidate $c$ has an
additional vote in group $G_B$, which can only be in subelection
$(C,V_1)$, as otherwise $c$ would be the unique winner of
$(C,V_2)$. Since in group $G_B$ candidate $e$ beats $p$ by $n-m-1$
points, there have to be at least $m$ additional groups from $G_i$ in
subelection $(C,V_1)$ (this is the only way $p$ can gain more points
than $e$). On the other hand, there can be at most $m$ groups from
$G_i$ in subelection $(C,V_1)$, as otherwise there would exist at least
one candidate in $B$ who would beat $p$ there. Thus 
there must be exactly 
$m$ groups from $G_i$ in subelection $(C,V_1)$, and these
groups have to correspond to an exact cover of $B$, since otherwise $p$
cannot beat all the $b_j$'s.~\end{proofs}

\end{document}